\documentclass[12pt]{article}
\input epsf      
\usepackage{graphicx}
\usepackage{amssymb}
\usepackage{amsfonts}
\usepackage{amsthm}
\usepackage{amsfonts}

\newcommand{\B}{{\mathbb B}}

\newcommand{\X}{{\mathbb X}}
\newcommand{\Y}{{\mathbb Y}}

\newcommand \ra {\rightarrow}

\newcommand{\ba}[1]{\begin{array}{#1}}
\newcommand{\ea}{\end{array}}

\newcommand{\be}{\begin{equation}}
\newcommand{\ee}{\end{equation}}
\newcommand{\bea}{\begin{eqnarray}}
\newcommand{\eea}{\end{eqnarray}}
\newcommand{\beann}{\begin{eqnarray*}}
\newcommand{\eeann}{\end{eqnarray*}}

\newcommand{\spin}{\sigma}
\newcommand{\spinb}{\bar{\sigma}}
\newcommand{\gas}{{n}}
\newcommand{\gasb}{\bar{n}}

\newcommand{\hb}{\bar{H}}

\newcommand{\size}{S}

\newcommand{\cf}{C_f}

\newcommand{\ch}{C_{\bar{H}}}

\newcommand{\cb}{C_B}

\def\reff#1{(\ref{#1})}
\begin{document}

\bibstyle{ams}

\title{Renormalization group maps for Ising models in lattice-gas variables}

\author{Tom Kennedy \\
Department of Mathematics \\
University of Arizona \\
Tucson, AZ 85721 \\
http://www.math.arizona.edu/$\, \hbox{}_{\widetilde{}}$ tgk \\
email: tgk@math.arizona.edu \\
}

\maketitle

\begin{abstract}
Real-space renormalization group maps, e.g., the majority rule 
transformation, map Ising-type models to Ising-type 
models on a coarser lattice. 
We show that each coefficient in the renormalized Hamiltonian
in the lattice-gas variables depends on only a finite number of 
values of the renormalized Hamiltonian. We introduce a method 
which computes the values of the renormalized Hamiltonian 
with high accuracy and so computes the coefficients in the 
lattice-gas variables with high accuracy. For the critical nearest 
neighbor Ising model on the square lattice with the majority rule 
transformation, we compute over 
1,000 different coefficients in the lattice-gas variable representation 
of the renormalized Hamiltonian  
and study the decay of these coefficients. We find that they decay
exponentially in some sense but with a slow decay rate. 
We also show that the coefficients in the spin variables are 
sensitive to the truncation method used to compute them. 
\end{abstract}

\newpage

\section{Introduction}

Real-space renormalization group transformations were introduced
to study critical behavior in Ising-type models. 
There has been extensive numerical study of these transformations,
and there is a rich picture of how they are believed to behave. 
However, there are essentially no mathematical results on these
transformations. 
The usual definition of these transformations is only formal 
since it involves an infinite-volume limit which must be proved to exist. 
The mathematical problem is to show that these renormalization 
group maps are rigorously defined in a neighborhood of the 
critical point, and to use them to study the  system in a neighborhood 
of the critical point. 
This is a difficult problem and the amount of rigorous progress that 
has been made is embarrassing. Starting with the critical nearest 
neighbor Hamiltonian, the first step of the renormalization group 
transformation has been proved to be defined for a few specific 
lattices and transformations \cite{tk_rga,tk_rgb}. 
The existence of the transformation well inside the high-temperature 
phase has been proved by rigorous expansion methods 
\cite{gp,israel,kash}. 
It is possible to construct examples of transformations for which 
the renormalized Hamiltonian can be proved to be non-Gibbsian,
including examples which start from the critical nearest neighbor
Ising model \cite{vefs,ve}. 

Even if we start with a finite-range Hamiltonian,
after just one step of the renormalization group 
transformation the renormalized 
Hamiltonian will be infinite range and have infinitely many 
different terms. The conventional wisdom is that 
they should decay both as the number of sites involved
grows and as the distance between these sites grows, so that the  
renormalized Hamiltonian may be well approximated by 
a finite number of terms. In some sense, this property is the 
raison d'\^etre of the renormalization group. 
It should allow one to study critical 
phenomena, which are inherently multiscale and so impossible to 
approximate well by a finite sets of terms, by studying a map 
of Hamiltonians which can be well approximated by a finite number of terms. 

Swendsen showed that one can compute the linearization of the 
renormalization group transformation about the fixed point from 
correlation functions that involve the original spins and the block spins
\cite{swendsen}. 
His method allows one to avoid computing any renormalized Hamiltonians. 
From the point of view of using the renormalization group to calculate
the critical exponents, this was a tremendous advance and was used in
a large number of subsequent Monte Carlo studies of the renormalization
group. From the point of view of trying to learn more about the 
renormalized Hamiltonians and the fixed point of the transformation, 
it had the unfortunate side effect that many of these Monte Carlo studies
did not compute any renormalized Hamiltonians.
In recent years there have been more studies that compute the 
renormalized Hamiltonian. In particular the 
the Brandt-Ron representation introduced in \cite{br} and 
studied further in \cite{ron_swen_a,ron_swen_b,ron_swen_br_a,ron_swen_br_b}
is similar to the method we use in this paper. 

The goals of this paper are to give a highly accurate 
method for computing the renormalized Hamiltonian which works in the 
lattice-gas representation and to use it to test
the conventional wisdom that the renormalization group transformation is 
well approximated by a finite number of terms. 
Our numerical calculations are done for the critical nearest neighbor 
Ising model on the square lattice, and we only consider the 
renormalized Hamiltonian obtained by a single application of the 
majority rule transformation using 2 by 2 blocks. 
However, our approach is quite general and can be applied to other 
dimensions, lattices and choices of the real-space renormalization
group transformation.

One of the key tenets of the renormalization group  is that
if we fix a block-spin configuration and study the original 
system subject to the constraint imposed by the block spins, then 
this constrained system is in a high-temperature phase even if the 
unconstrained system is at its critical point. As an extreme 
case consider the block-spin configuration of all $+1$'s with the 
majority rule transformation. The effect of this constraint on the 
original Ising system is similar to imposing a positive magnetic 
field, and the constrained system should have a relatively short 
correlation length.  Our computational method for the 
renormalization group transformation 
takes advantage of this property.

In the next section we review the definition of real-space 
renormalization group transformations. 
In section three we explain our method for computing the renormalized 
Hamiltonian in the lattice-gas representation. Some of the details
are postponed to section five. We use this method to
study the decay of the terms in the renormalized Hamiltonian. 
In section four we consider how to compute the renormalized 
Hamiltonian in the more standard spin variables. There are multiple 
ways to do this, and we will see that the computed value of an individual
coupling coefficient in the renormalized Hamiltonian varies 
considerable with the method used. 
We also study the decay of 
the renormalized Hamiltonian in the spin variables. 
Section five provides further detail for our method for 
computing the renormalized Hamiltonian. We consider the 
various sources of error in our computations in section six, and offer
some conclusions in section seven. 

The significant dependence of coefficients in the renormalized 
Hamiltonian on the truncation method used has been seen before. 
In particular, Ron and Swendsen observed a change of several percent
in the nearest neighbor coupling when the number of couplings kept 
was changed from six to twelve \cite{ron_swen_a}.
In \cite{ron_swen_b} they wrote 
``Even though the individual multispin interactions usually
have smaller coupling constants than two-spin interactions,
the fact that they are very numerous can lead to multispin 
interactions dominating the effects of two-spin interactions.''
Truncating the space of Hamiltonians implies that the linearization 
of the renormalization group map about the fixed point is also truncated.
An early, interesting paper on the effect of this truncation is 
\cite{sgm}.

\section{Real-space renormalization group transformations}

In this section we quickly review the definition of real-space
renormalization group transformations. We refer the reader to 
\cite{vefs} for more detail. 

\begin{figure}[tbh]
\begin{center}
\includegraphics{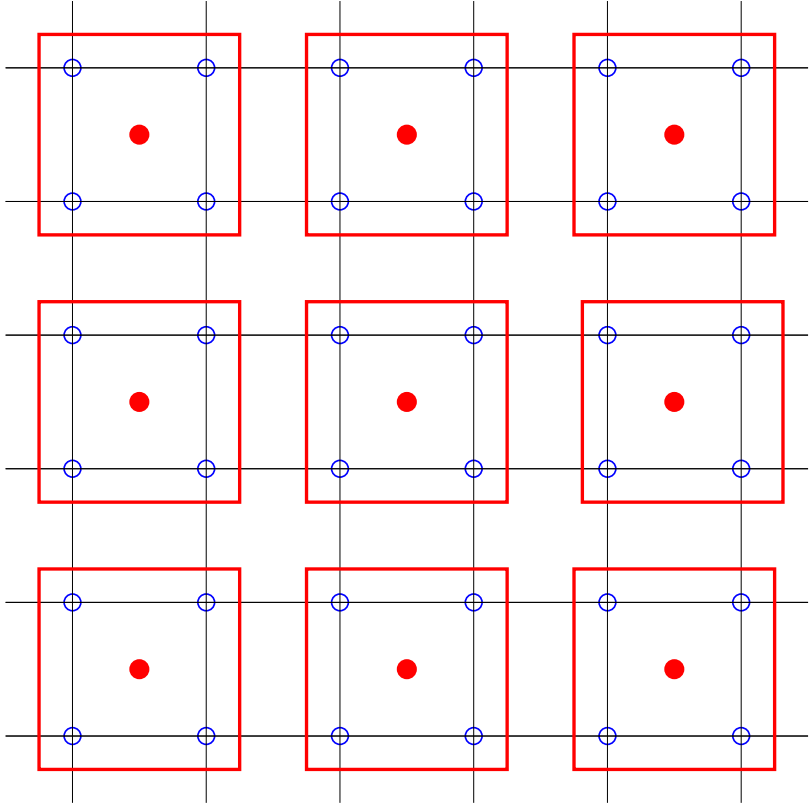}
\caption{\leftskip=25 pt \rightskip= 25 pt 
A renormalization group blocking for the square lattice. 
The original lattice sites are the open circles, the sites after 
the renormalization transformation are the solid circles.
}
\label{sq_rg}
\end{center}
\end{figure}

Consider an Ising-type model in which the spins take on only the values
$\pm 1$. The lattice is divided into blocks and each block is assigned
a new spin variable called a block spin. The example of the square 
lattice with $2$ by $2$ blocks is shown in figure \ref{sq_rg}.
We consider transformations
in which the block spins also take on only the values $\pm 1$. 
The transformation is specified by a kernel $T(\spinb,\spin)$.
Here $\spin$ denotes the original spins and $\spinb$ the block 
spins. The kernel is required to satisfy
\bea
\sum_{\spinb} \, T(\spinb,\spin) = 1
\label{kcond}
\eea
for all original spin configurations $\spin$. 
The renormalized Hamiltonian $\hb(\spinb)$
is formally defined by 
\bea
 e^{- \hb(\spinb) } = \sum_{\spin}
\, T(\spinb,\spin) \, e^{-H(\spin)}
\label{rgdef}
\eea
(Note that the inverse temperature $\beta$ has been absorbed into the 
Hamiltonians in the above equation.) 
This is only a formal definition since we must first restrict to a finite
volume in order to make sense of this equation. Proving that 
the finite-volume definition of $\hb$ has an infinite-volume limit
is essentially an open problem. 
The condition \reff{kcond} implies that 
\bea
\sum_{\spinb} e^{- \hb(\spinb) } = \sum_{\spin} 
\, e^{-H(\spin)}
\nonumber
\eea
so that the free energy of the original model can be recovered
from the renormalized Hamiltonian. This property allows one to 
study the critical behavior of the system by studying iterations 
of the renormalization group map. In particular, 
the critical exponents may be related to the eigenvalues of 
the linearization of the map about its fixed point.

One widely studied family of kernels is the family of 
majority rule transformations.
If there are an odd number of spins in every block, then 
$T(\spinb,\spin)=1$ if the majority of the spins in each 
block agree with the block spin and $T(\spinb,\spin)=0$ otherwise. 
If there are an even number of spins in every block, then we let 
$T(\spinb,\spin)$ be the product over the blocks $B$ of 
\bea
t(\spinb_B,\{\spin_i\}_{i\in B})=
 \cases{ 1 & if \, $\spinb_B \, \sum_{i \in B} \spin_i >0$ \cr
         0 & if \, $\spinb_B \, \sum_{i \in B} \spin_i <0$ \cr
         1/2 & if \, $\sum_{i \in B} \spin_i =0$ \cr}
\eea
where $\spinb_B$ denotes the block spin for block $B$.

The general approach presented in this paper applies to 
all these renormalization group maps. The numerical 
calculations that we will present are for 
the critical nearest neighbor Ising model on the square lattice 
with the majority rule renormalization group map with two by two blocks.

\section{Renormalized Hamiltonian in the lattice-gas variables}
\label{gas_sect}

Real-space renormalization group calculations are usually done 
using the spin variables $\spin_i = \pm 1$. Our method is based on 
what are sometimes called the lattice-gas variables 
$\gas_i =(1-\spin_i)/2$ which take on the values $0,1$. 
Note that we have made the convention that a spin value of $+1$ corresponds
to a lattice gas value of $0$. 
Throughout this paper we will use $\spin$'s for spin variables taking 
on the values $\pm 1$, and $\gas$'s for lattice-gas variables taking 
on the values $0,1$. 
We indicate renormalized spins or variables with a bar over them, 
e.g., $\spinb_i$, $\gasb_i$. We use $\spin$ to denote the 
entire spin configuration $\{ \spin_i\}$. Likewise, $\gas$, $\spinb$ and 
$\gasb$ denote the corresponding collections of variables.

In this section we work entirely in the lattice-gas variables, both 
for the original Hamiltonian and the renormalized Hamiltonian. 
We write the renormalized Hamiltonian as 
\bea
\hb(\gasb) = \sum_Y \, c(Y) \, \gasb(Y) 
\label{gas_coefs}
\eea
where the sum is over all finite subsets including the empty set, and 
\beann
\gasb(Y) = \prod_{i \in Y} \gasb_i
\eeann

Consider the block-variable configuration of all $0$'s. 
Our method for computing the renormalized Hamiltonian uses only 
block-variable configurations which differ from this configuration at a
finite number of sites. 
For a finite subset $X$, let $\gasb^X$ denote the block-variable 
configuration with all block variables in $X$ equal 
to $1$ and the rest equal to $0$. Then eq. \reff{rgdef} says 
\beann
\exp(-\hb(\gasb^X))  
 = \sum_\gas \, T(\gasb^X,\gas) \, e^{-H(\gas)}
\eeann
Note that $\gasb^\emptyset(X)=0$ except when $X=\emptyset$. 
So $\hb(\gasb^\emptyset) = c(\emptyset)$. 
In particular, $c(\emptyset)$ will 
grow as the size of the finite volume. The other coefficients 
$c(Y)$ should have finite limits in the infinite-volume limit. 
We define $f(X)$ by 
\bea
f(X) = \hb(\gasb^X) - \hb(\gasb^\emptyset) 
\nonumber
\eea
Then $f(X)$ should have a finite limit in the infinite-volume limit,
and it should be related to the infinite-volume $c(X)$ by 
\bea
f(X) = \sum_{Y: \emptyset \ne Y \subset X} c(Y)
\label{free_eq}
\eea


The system of equations \reff{free_eq} can be explicitly solved for the $c(Y)$.
We claim that the solution for $X \ne \emptyset$ is  
\bea
c(X) = \sum_{Y:\emptyset \ne Y \subset X} (-1)^{|X|-|Y|} \, f(Y)
\label{coef_sol}
\eea
This is a standard inversion trick. To verify \reff{coef_sol}, 
define $c(X)$ by \reff{coef_sol}.
Then for a given $X \ne \emptyset$, 
\bea
\sum_{Y:\emptyset \ne Y \subset X} c(Y) &=& \sum_{Y:\emptyset \ne Y \subset X} 
\sum_{Z:\emptyset \ne Z \subset Y} (-1)^{|Y|-|Z|} \, f(Z)
\nonumber \\
&=& \sum_{Z:\emptyset \ne Z \subset X} \, f(Z) 
\sum_{Y:Z \subset Y \subset X} (-1)^{|Y|-|Z|} 
\label{check}
\eea
The sum over $Y$ is $1$ if $X=Z$. If $Z$ is a proper subset of $X$, we 
claim this sum is $0$. To see this:
\beann
\sum_{Y:Z \subset Y \subset X} (-1)^{|Y|-|Z|} 
= \sum_{W:W \subset X \setminus Z} (-1)^{|W|}
= \prod_{i \in  X \setminus Z} (1 + (-1)) =0
\eeann
Thus \reff{check} collapses to $f(X)$. 

The important feature of eq. \reff{coef_sol} is that the coefficient
$c(X)$ only depends on a finite number of free energies $f(Y)$, 
specifically those with $Y \subset X$. As we will see, these free 
energies can be computed extremely accurately. So individual
coefficients $c(X)$ in the lattice-gas variables can  be computed
extremely accurately. Moreover, this computation does not depend on 
how many terms we decide to keep in the renormalized Hamiltonian. 
If we increase the number of terms we keep, then the coefficients 
we have already computed will not change. 

We have carried out numerical calculations of a large number of
the coefficients in the lattice-gas representation for the 
critical nearest neighbor Ising model on the square lattice 
with the majority rule renormalization group map with two by two blocks. 
We need a criterion for deciding for which $Y$ we will compute $c(Y)$.
We assume the coefficients will decay as 
the number of sites in $Y$ grows and as the distance between these 
sites grows. So we need a measure of the size of a set $Y$. 
There is no canonical way to define this size.
We use the following ad hoc quantity.
If $Y=\{y_1,y_2,\cdots,y_n\}$, then we define 
\bea
\size(Y) = \sum_{i=1}^n ||y_i - c||_2^2
\label{size_def}
\eea
where $c$ is the center of mass:
\bea
c= {1 \over n} \sum_{i=1}^n y_i
\nonumber 
\eea
and $||\quad||_2$ is the usual Euclidean distance in the plane. 
Note that we do not take a square root in \reff{size_def}.

We claim that if $X^\prime \subset X$ then $\size(X^\prime) \le \size(X)$.
To prove this it suffices to prove the case that $X^\prime$ has 
one less site than $X$. Let $X$ be $x_1,x_2,\cdots,x_n$ and $X^\prime$ be
$x_1,x_2,\cdots,x_{n-1}$. Let $c$ be the center of mass of $X$ and 
$c^\prime$ the center of mass of $X^\prime$. 
So 
\bea
\size(X^\prime) = \sum_{i=1}^{n-1} ||x_i - c^\prime||_2^2
\nonumber 
\eea
The center of mass has the property that it minimizes the 
above sum. So 
\bea
\size(X^\prime) \le \sum_{i=1}^{n-1} ||x_i - c||_2^2
\le \sum_{i=1}^n ||x_i - c||_2^2 = \size(X)
\nonumber 
\eea

We fix a cutoff $C>0$, and compute $c(Y)$ for all $Y$ with
$\size(Y) \le C$. We only need to compute it for one $Y$ from each 
translation class, and so there are a finite number of such $Y$'s. 
The bulk of the computation is computing the free energies $f(X)$ 
for $X$ with $\size(X) \le C$. Using eq. \reff{coef_sol}
to find the $c(Y)$ requires comparatively little computation.
The property that $X^\prime \subset X \Rightarrow
\size(X^\prime) \le \size(X)$, 
implies that the collection of $X$ for which we must compute $f(X)$ 
is just the collection of $X$ with $\size(X) \le C$. 

\begin{figure}[tbh]
\begin{center}
\includegraphics{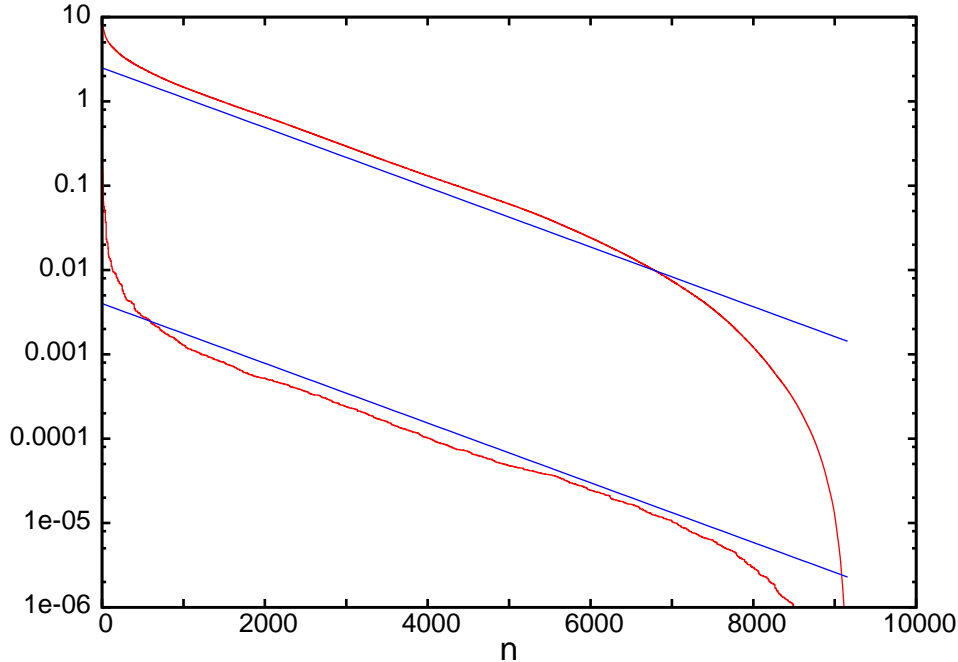}
\caption{\leftskip=25 pt \rightskip= 25 pt 
The coefficients $c(Y_n)$ are ordered so $|c(Y_n)|$ decreases. 
The bottom curve is $|c(Y_n)|$ vs. $n$, and the top curve is 
the tail $\sum_{i=n}^N |c(Y_n)|$ vs. $n$.  
}
\label{zero_one_decay}
\end{center}
\end{figure}

To study how fast the coefficients $c(Y)$ decay, we take one 
coefficient from each translation class that we have 
computed and order them so that $|c(Y)|$ is decreasing, i.e., 
$|c(Y_n)| \ge |c(Y_{n+1})|$.
We then plot $|c(Y_n)|$ as a function of $n$.
This is the bottom curve in figure \ref{zero_one_decay}. 
Note that the vertical axis uses a logarithmic scale. 
The second quantity plotted (the top curve in the figure) is 
\beann
\sum_{i=n}^N |c(Y_i)|
\eeann
as a function of $n$, where $N$
is the total number of $Y$ for which we compute the coefficients.
The two lines shown are given by $c 2^{-n/850}$ for two different
values of $c$. 
The two curves in the figure depend on the function $\size(Y)$ we use to 
measure the size of sets and the cutoff we use for this function. 
However, whatever function and cutoff we use, the resulting curve 
will be a lower bound on the curve that would result from 
computing all the coefficients $c(Y)$. 
In particular, we can make the following observations. 
The lower curve crosses the horizontal lines at $10^{-2}$, $10^{-3}$ and 
$10^{-4}$ at $131$,$1223$ and $4023$, respectively. 
Hence there are at least $131$ translation classes with a coefficient
bigger than $10^{-2}$, at least $1223$ with a coefficient
bigger than $10^{-3}$, and at least $4023$ with a coefficient
bigger than $10^{-4}$.

In the preceding discussion we used one coefficient from each 
translation class. In addition to the translation symmetry the 
model is also symmetric under rotations by 90 degrees and relections 
in lattice axes. More precisely, the additional symmetry is the 
dihedral group of order 8. We have repeated the previous study of 
the decay of the coefficients taking into account the dihedral 
group symmetry as well as the translational symmetry by taking 
only one term from the above list from each dihedral group symmetry 
class. The main effect is that the scale on the horizontal 
axis is reduced by a factor of $8$. 
This is not surprising since for most subsets, rotations and reflections
will generate eight different subsets. 

From a mathematical perspective, one would like to show that the 
renormalized Hamiltonian exists in some Banach space. 
One choice of norm for the Banach space would be 
\bea
\sum_{Y : 0 \in Y} |c(Y)|
\nonumber 
\eea
One would like to approximate the Hamiltonian by a Hamiltonian 
with a finite number of terms. So it is important to see how 
fast the above sum converges as we include more terms in the 
Hamiltonian. This is similar to the second plot in figure 
\ref{zero_one_decay}. Note that in this norm each translation
class appears $|Y|$ times. So the  second plot in figure 
\ref{zero_one_decay} is in some sense a lower bound on the decay 
for the Hamiltonian. 
Another choice of the norm would be 
\bea
\sum_{Y : 0 \in Y} |c(Y)| e^{\mu(Y)}
\nonumber 
\eea
for some measure $\mu(Y) \ge 0$ of the size of $Y$. For this norm
the convergence would be even slower than that seen in the figure.

It is worth noting that norms defined using the lattice-gas 
representation of the Hamiltonian are in general stronger than norms that 
use the spin variable representation \cite{israel_book}. 
For example, if we can write the 
Hamiltonian in the lattice-gas representation \reff{gas_coefs} with 
\beann
\sum_{Y \ni 0} \, |c(Y)| <\infty
\eeann
then it is straightforward to show that 
the Hamiltonian can be represented in the spin variable representation
\reff{spin_coefs} with 
\beann
\sum_{Y \ni 0} \, |d(Y)| <\infty
\eeann

\section{Renormalized Hamiltonian in the spin variables}

In the previous section we saw that in the lattice-gas variables
there is a natural way to compute the coefficients $c(Y)$ in 
the expansion \reff{gas_coefs} for $\hb$. 
In this section we consider the renormalized Hamiltonian in 
the spin variables:
\bea
\hb(\spinb) = \sum_Y \, d(Y) \spinb(Y) 
\label{spin_coefs} 
\eea
with $\spinb(Y) = \prod_{i \in Y} \spinb_i$. 
The sum over $Y$ is over all finite subsets. 

We can use $\gasb_i=(1-\spinb_i)/2$ to express the spin coefficients
$d(Y)$ in \reff{spin_coefs} in terms of the lattice-gas coefficients
$c(Y)$ in \reff{gas_coefs}. 
\bea
\hb(\spinb) &=& \sum_X c(X) \gasb(X) 
= \sum_X c(X) \, 2^{-|X|} \prod_{i \in X} (1-\spinb_i)  
\nonumber \\
&=& \sum_X c(X) \, 2^{-|X|} \sum_{Y: Y \subset X} (-1)^{|Y|} \spinb(Y)
\nonumber \\
&=& \sum_Y \spinb(Y) (-1)^{|Y|} \sum_{X: Y \subset X}  c(X) \, 2^{-|X|} 
\nonumber \\
&=& \sum_Y d(Y) \, \spinb(Y) 
\nonumber
\eea
where the spin coefficients $d(Y)$ are given by 
\bea
d(Y) = (-1)^{|Y|} \sum_{X: Y \subset X}  c(X) \, 2^{-|X|} 
\label{exact_d}
\eea
The problem is that to compute the spin coefficient $d(Y)$ we need 
the lattice-gas coefficients $c(X)$ for infinitely many $X$, and so 
we need the free energies $f(X)$ for infinitely many $X$'s. 
So we must  introduce some sort of approximation.

Let $\Y_\infty$ be a collection of finite subsets of the renormalized lattice
such that one set from each translation class is 
contained in $\Y_\infty$. We can rewrite \reff{spin_coefs} as
\beann
\hb(\spinb) = \sum_{Y \in \Y_\infty} \, d(Y) \sum_t \spinb(Y+t) 
\eeann
where the sum over $t$ is over the translations for the renormalized 
lattice. Here $Y+t$ denotes $\{i+t: i \in Y\}$. 

Now let $\Y$ be a {\it finite} subcollection of $\Y_\infty$.
We want to compute an approximation to the above 
of the form
\bea
\hb(\spinb) \approx \sum_{Y \in \Y} \, d(Y) \sum_t \spinb(Y+t) 
\nonumber
\eea
We will consider two methods which we will refer to as the ``partially 
exact'' method and the ``uniformly close'' method. 

For the partially exact method, we begin by noting that we can
write $\hb(\gasb)$ as 
\beann
\hb(\gasb) = \sum_{Y \in \Y_\infty} \, c(Y) \sum_t \gasb(Y+t) 
\eeann
The approximation is simply to truncate this sum by 
restricting $Y$ to those in $\Y$:
\beann
\hb(\gasb) \approx \sum_{Y \in \Y} \, c(Y) \sum_t \gasb(Y+t) 
\eeann
The $c(Y)$ are exact. As we saw in the last section we can 
compute them from \reff{coef_sol} by computing the free energies $f(X)$
for $X \in \Y$.
We then convert this Hamiltonian to the spin variables with no approximation. 
The result is that the approximation to $\hb(\spinb)$ is 
\bea
\sum_{Y \in \Y} \, d(Y) \sum_t \spinb(Y+t) 
\eea
with 
\bea
d(Y) = (-1)^{|Y|} \sum_{X : Y \subset X, X+t \in \Y}  c(X) \, 2^{-|X|} 
\eea
where the notation $X+t \in \Y$ means some translation of $X$
(possibly $X$ itself) is in $\Y$.
Thus this method is equivalent to truncating the exact formula
\reff{exact_d} by restricting the sum over $X$ to sets in $\Y$
and their translates. 
In the lattice-gas variables our approximation to $\hb$ agrees
with the true $\hb$ for all $n^X$ such that $X \in \Y$. 
The change from lattice gas to spin variables did not 
involve any approximation, so our 
approximation to $\hb$ in the spin variables agrees exactly with the true
$\bar{H}$ for all configurations $\spinb^Y$ for $Y \in \Y$.
This is the reason for calling this method ``partially exact.''
It is exact for some of the block-spin configurations. 

For the uniformly close method let $\X$ be another finite collection of 
finite subsets which contains at most one set from each translation
class. We compute the free energies $f(X)$ for $X \in \X$, i.e., 
we compute $\hb(\spinb^X)$.  
We define the error of a set of coefficients $\{d(Y): Y \in \Y \}$
to be 
\beann
\max_{X \in \X} | \hb(\spinb^X) - 
  \sum_{Y \in \Y} \, d(Y) \sum_t \spinb^X(Y+t) |
\eeann 
where $\spinb^X$ is the spin configuration which is 
$-1$ on $X$ and $+1$ on all other sites. 
We then choose the coefficients $d(Y)$ to 
minimize the above error. This is a standard linear programming problem
which we solve by the simplex algorithm. 
We call this the uniformly close approximation since we have a uniform
bound on the difference between our approximation and the exact $\hb$ 
for the block-spin configurations $\spinb^X$ for $X \in \X$.
(For other $X$ we cannot say anything about how well the approximation
does.) If $\X=\Y$, then the partially exact approximation makes the 
above error zero. We only use the uniformly close approximation 
for $\X$ which are larger than $\Y$.

We take the following point of view. We think of $\Y$ as being fixed. 
It determines a finite-dimensional space of Hamiltonians that 
we use to approximate
the renormalized Hamiltonian. We then think of the collection $\X$ 
as being variable. A larger $\X$ means we ``know'' more free energies 
and so have more information to use in computing the approximation. 
In our studies we will take the collection $\Y$ to be all the 
subsets $Y$ with $s(Y) \le \ch$ for some cutoff $\ch$,  
and $\X$ to be all the $X$ with $s(X) \le \cf$ for some cutoff
$\cf \ge \ch$   

When we worked in the lattice-gas variables the computation of 
the coefficients $c(X)$ was unambiguous. The computation 
of the values of $\hb(\gasb)$ requires some approximations, 
but as we will see in section \ref{compute_free}
these approximation are well behaved and introduce 
small errors. The computation of the $c(X)$ from the 
$\hb(\gasb)$ does not require any approximation or truncation.
In the spin variable representation we now have multiple ways to 
compute the coefficients $d(X)$ depending on whether we use
the partially exact or uniformly close methods and on the 
choices of $\X$ and $\Y$. We restrict our study of the spin 
variable coefficients to studying how these choices affect the 
values of individual coefficients. We focus our attention on 
three particular coefficients: the nearest neighbor, the next
nearest neighbor and the plaquette. These refer to the coefficients
of $\sigma_i \sigma_j$ with $|i-j|=1$,
of $\sigma_i \sigma_j$ with $|i-j|= \sqrt{2}$, and 
of $\sigma_i \sigma_j \sigma_k \sigma_l$ where $i,j,k,l$ are the 
corners of a unit square. 
As in the previous section, our numerical calculations are for
the critical nearest neighbor Ising model on the square lattice 
with the majority rule renormalization group map with two by two blocks. 

\begin{figure}[tbh]
\includegraphics{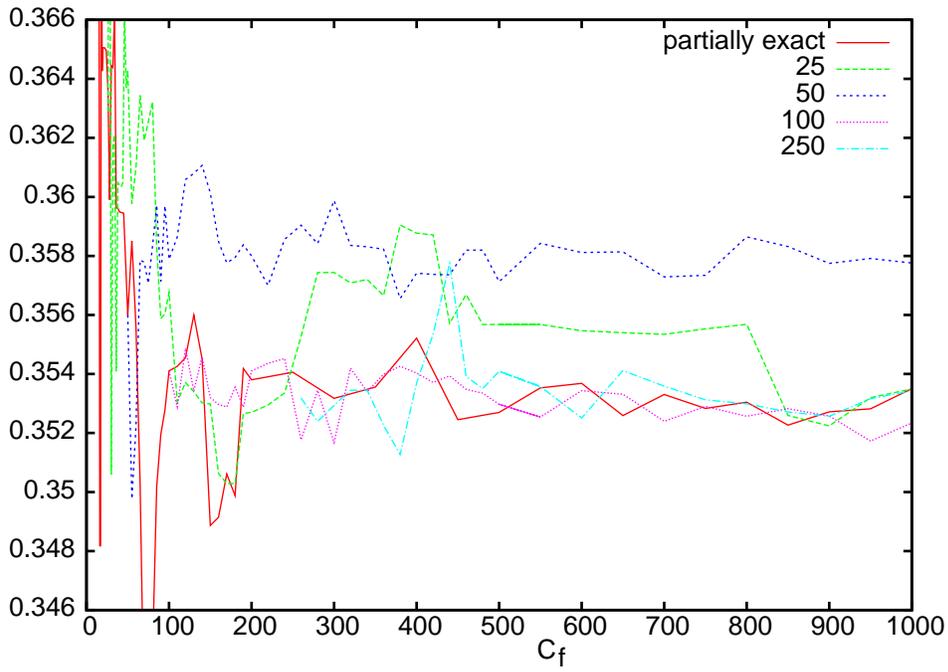}
\caption{\leftskip=25 pt \rightskip= 25 pt 
The dependence of the nearest neighbor coefficient in the spin variable 
representation of the renormalized Hamiltonian on the computation method.
The solid curve is the partially exact method. The dashed curves are the
uniformly close method with four different choices of $\ch$
}
\label{coef1}
\end{figure}

\begin{figure}[ht]
\begin{center}
\includegraphics{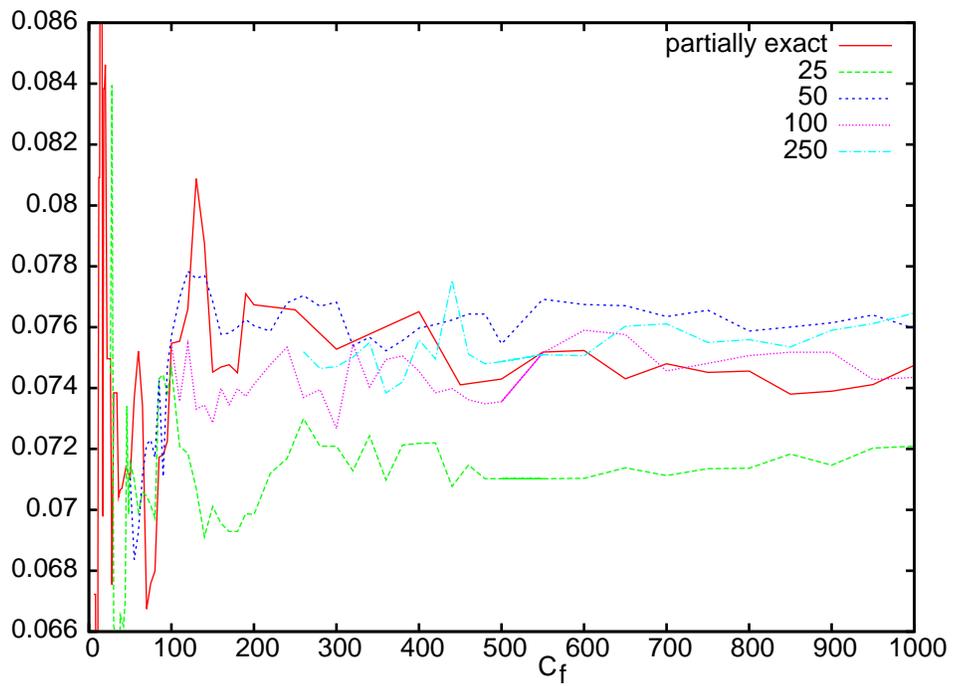}
\caption{\leftskip=25 pt \rightskip= 25 pt 
The next nearest neighbor coefficient in the spin variable 
representation of the renormalized Hamiltonian 
}
\label{coef2}
\end{center}
\end{figure}

\begin{figure}[ht]
\begin{center}
\includegraphics{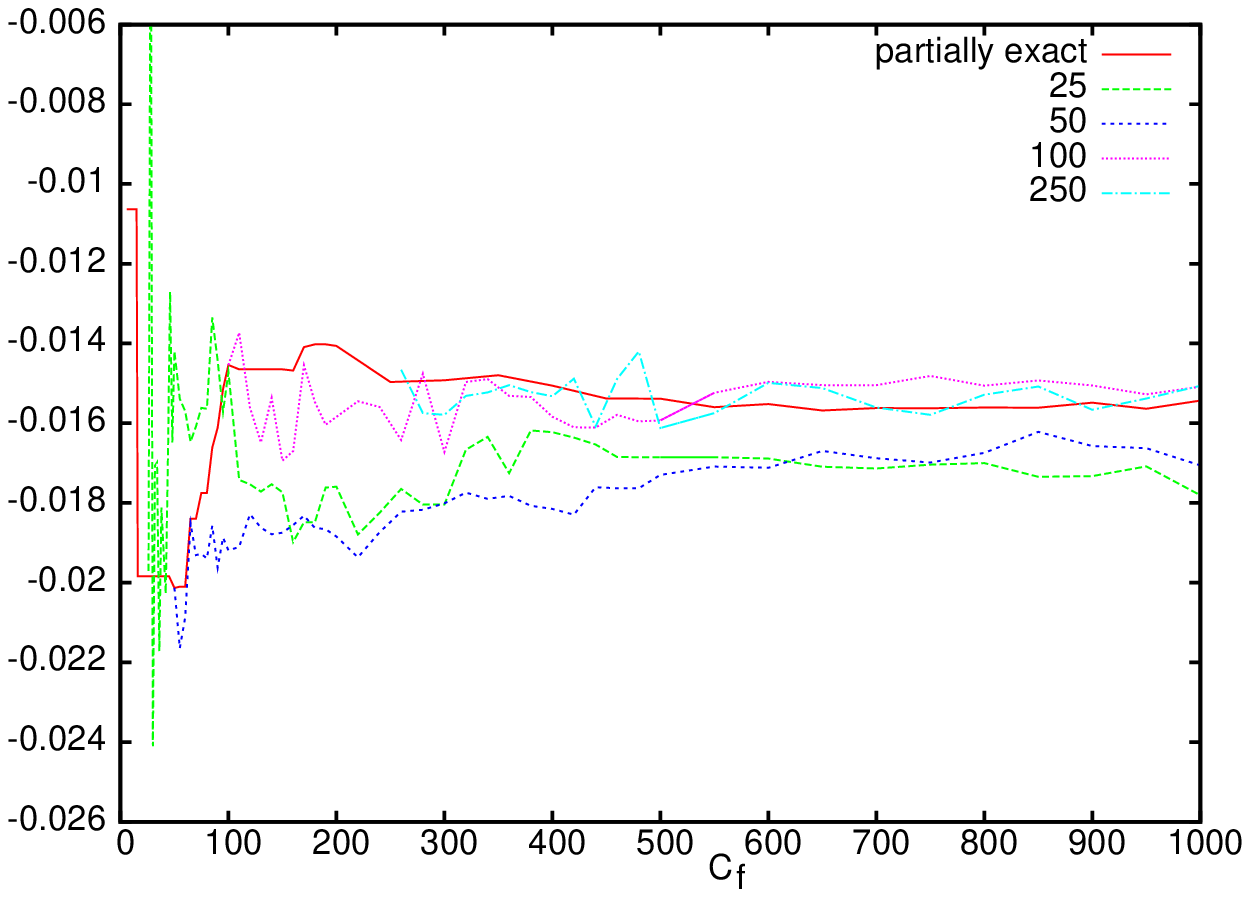}
\caption{\leftskip=25 pt \rightskip= 25 pt 
The plaquette coefficient in the spin variable 
}
\label{coef3}
\end{center}
\end{figure}

For the partially exact method we have one parameter - the cutoffs 
$\ch$ and $\cf$ are equal and correspond to the cutoff $C$ of 
section \ref{gas_sect}.
So we can plot individual coefficients as a function of $\cf$. 
For the uniformly close method we have two 
parameters: the cutoff $\ch$ determines the finite-dimensional 
subspace used to approximate the renormalized Hamiltonian and 
the cutoff $\cf$ determines the number of $\hb(\gasb)$ values 
we use. We plot the coefficients in this case as a function of 
$\cf$ for several different choices of $\ch$. 
The results are shown in figures \ref{coef1},\ref{coef2}, and 
\ref{coef3}. The variations seen in the three coefficients
are roughly comparable in size. 
Note that while the ranges of the vertical 
axes vary in the three figure, the scales for the vertical axes 
are all the same. The variations in the coefficients are 
on the order of several thousandths. 
So even for these relatively large coefficients, 
it is difficult to determine the value of the coefficient to 
better than a few percent.
For smaller coefficients the variations are somewhat smaller, 
but as a fraction of the coefficient they are typically larger. 

These three coefficients (along with many others) have of course 
been computed before. Two early references are
\cite{gc,swendsen_84}.
The point of our study is not the values of these coefficients but 
rather the variation in their values as one varies the 
method used to compute them. 

\clearpage

\section{Computing the free energy}
\label{compute_free}

Fix a block-spin configuration $\spinb$. We want to compute
the free energy $\hb(\spinb)$ of the constrained partition function
\beann
\exp(-\hb(\spinb)) = \sum_{\spin} \, T(\spinb,\spin) \, e^{-H(\spin)}
\eeann
Initially we work with the spin variables, but later in this section we
will switch to the lattice-gas variables. 
We only need to do this computation for configurations 
$\spinb$ which are $+1$ except on a finite set. 
Even when the original system is at the critical point, these 
constrained systems have relatively short correlation lengths. 
This is where the real power of the renormalization group becomes
apparent. In particular, finite-volume effects in the above 
computation decay very quickly as the volume increases. 

Before we explain our method for the computation, we first indicate 
how $\hb(\spinb)$ can be computed by a Monte Carlo calculation. 
(We have done such a simulation as a check on the method we describe 
later.) Fix a relatively small set of block spins, $X$. Let $V$ be a
finite volume of block spins containing $X$ which is large enough that 
the boundary of $V$ is far from $X$. 
We include only the factors in the renormalization 
group kernel corresponding to the blocks in $V \setminus B$. For these 
blocks we take the block spins to be $+1$. 
We then run a Monte Carlo simulation of the Ising system with this kernel
outside of $X$. When we sample the simulation we compute the 
block-spin configuration on $X$. This allows us to compute the relative 
weights of the possible block-spin configurations on $X$. 
From these weights we can then compute the $\hb(\spinb)$ for 
$\spinb$ which are $-1$ only on a subset of $X$. 

\begin{figure}[tbh]
\begin{center}
\includegraphics{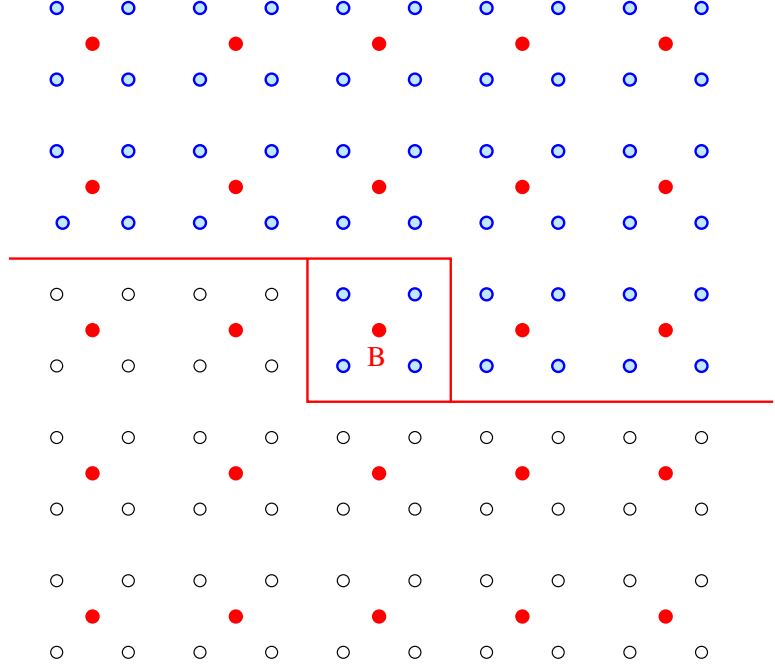}
\caption{\leftskip=25 pt \rightskip= 25 pt 
We compute $\hb(\spinb)$ by summing out the original spins 
one block at a time. The open circles are spins that have been summed
over, while the blue (gray) circles are spins that have yet to be summed over.
The red (black) circles are the fixed block spins.}
\label{fpe}
\end{center}
\end{figure}

We now turn to our method for computing $\hb(\spinb)$. 
It does not involve Monte Carlo methods, and it is much more accurate 
than the Monte Carlo approach described in the previous paragraph.
Everything in the following depends on $\spinb$, but we will 
not make this dependence explicit. 
Imagine that we have summed over the spins one block at a time 
in such a way that we have reached the state in figure \ref{fpe}.
Open circles indicate sites in the original lattice 
for which we have already summed over the spin, and blue (gray) circles
represent sites for which we have not. (Red (solid) circles 
indicate the block spins which are fixed throughout this
computation.) The result of this partial computation of the free energy 
is a function of the spins in the original lattice with shaded circles.
In fact, it only depends on those that are 
nearest neighbors of a spin with an open circle. We will 
refer to these spins as boundary spins. The quantity we have 
computed so far is positive, and we write it in the form
\beann
\exp(\sum_X a(X) \spin(X) + \Delta H) 
\, \prod_{B^\prime}  t_{B^\prime}(\spinb,\spin)
\eeann
where $X$ is summed over finite subsets of the boundary spins. 
$\Delta H$ denotes the terms in the Hamiltonian that only involve 
spins with shaded circles. (These terms have not yet entered 
the computation.) The product over $B^\prime$ is over the blocks 
containing shaded circles, and $t_{B^\prime}(\spinb,\spin)$ 
is the factor in the renormalization group kernel for block $B^\prime$. 
The next step is to sum over the four spins in the block $B$ and 
take the logarithm of the result:
\bea
\ln \left[ \sum_{\spin_B} \exp(\sum_X a(X) \spin(X) + \Delta H)
\, \prod_{B^\prime}  t_{B^\prime}(\spinb,\spin)
\right]
\nonumber
\eea
The sum over $\spin_B$ denotes a sum over the spins $\spin_i$ with 
$i \in B$. 
Terms $a(X) \spin(X)$ for which $X \cap B = \emptyset$ pass through
this computation trivially. So do the terms in $\Delta H$ 
which do not involve a spin in the block $B$ and the factors 
$t_{B^\prime}$ for $B^\prime \ne B$. So the 
computation that we must actually do is 
\bea
\ln \left[ \sum_{\spin_B} \exp(\sum_{X: X \cap B \ne \emptyset}
 a(X) \spin(X) + h) \, t_B(\spinb,\spin) \right]
\nonumber
\eea
where $h$ contains the terms in $H$ that only depend on spins 
with shaded circles and depend on at least one spin in $B$. 

To do this computation numerically, we must introduce a truncation. 
We fix a finite subset $D$ of the boundary sites centered near $B$. 
We then restrict the sum over $X$ to $X \subset D$. We need 
to write the result of the truncated computation in the form 
\bea
\ln \left[ \sum_{\spin_B} 
\exp(\sum_{X: X \cap B \ne \emptyset, X \subset D}
 a(X) \spin(X) + h) \, t_B(\spinb,\spin) \right]
= \sum_Y a^\prime(Y) \spin(Y) 
\nonumber
\eea
The left side only depends on spins in $D^\prime=D \setminus B$, 
so the sum on the right may be restricted to $Y \subset D^\prime$.
If we define $F(\spin)$ to be the left side of this equation, then 
the coefficients are given by 
\bea
a^\prime(Y) = 2^{-|D^\prime|} \sum_{\spin_{D^\prime}} F(\spin) 
\nonumber
\eea
The amount of computation required grows quite rapidly as $D$ 
grows for three reasons. First, the number of $X$ with $X \subset D$ 
grows as $2^{|D|}$. Second, the sum over $\spin_{D^\prime}$ in the above 
also grows as  $2^{|D|}$. Third, the number of $Y$ also grows as 
$2^{|D|}$. We have found that $a(X)$ decays 
quickly as the number of sites in $X$ grows. So we can make a 
further truncation by only keeping terms $a(X)$ with $|X|$ 
less than some specified cutoff. 
($|X|$ denotes the number of sites in $X$.)
This greatly reduces the growth of 
the computation with $D$ from the first and third effects. But 
we are still left with the second effect. 

We can eliminate the second effect by working in the lattice-gas 
variables. We replace $\sum_X a(X) \spin(X)$ by $\sum_X b(X) \gas(X)$.
Define 
\bea
F(\gas) = \ln \left[ \sum_{\gas_B} 
\exp(\sum_{X: X \cap B \ne \emptyset, X \subset D}
 b(X) \gas(X) + h) \, t_B(\gasb,\gas) \right]
\eea
We need to compute the coefficients in 
\bea 
F(\gas) = \sum_X b^\prime(X) \gas(X)
\nonumber
\eea
As we saw in section \ref{gas_sect}, they are given by 
\bea
b^\prime(X) = \sum_{Y:\emptyset \ne Y \subset X} (-1)^{|X|-|Y|} \, F(n^Y)
\label{bprime}
\eea
where $n^Y$ is the configuration that is $1$ on $Y$ and $0$ off of it. 
So to compute $b^\prime(X)$ we only need to compute $F(n^Y)$ for 
$Y \subset X$. 

In this approach using the lattice-gas variables we can forget about 
the set $D$ entirely. Instead we specify a finite collection $\B$ 
of subsets of the boundary spins with the property that they 
intersect $B$. We then make the approximation
\bea
\sum_{X: X \cap B \ne \emptyset}  b(X) \gas(X)  \approx
\sum_{X \in \B: X \cap B \ne \emptyset}  b(X) \gas(X)  
\eea
We use \reff{bprime} to compute $b^\prime(X)$.
It will be nonzero only for $X \subset D^\prime$. 
Before we sum over the next block of spins, we need to truncate 
$\sum_X b^\prime(X) \gas(X)$. We keep only the terms such that $X$ 
is in $\B+t$ where $t$ is the translation that takes the block we just 
summed over to the block we are summing over next, and $\B+t$ 
denotes the collection of sets of the form $X+t$ for $X \in \B$.  

We take the finite collection $\B$ to be all $X$ which intersect $B$ 
and satisfy $\size(X) \le \cb$ where $\size(X)$ is some size function 
and $\cb$ is some cutoff. 
We use the size function given by \reff{size_def}
that we used for choosing the block-spin sets. 
In our calculations we take 
$\cb=260$ which leads to $10,763$ sets in the collection $\B$. 
We discuss the effect of $\cb$ on the error in the following section.

The above discussion took place in an infinite volume. 
The region shown in figure \ref{fpe} is a finite piece of the infinite
volume.
In practice we can only sum over the spins in a finite number of blocks.
The block-spin configuration $\gasb$ is of the form $\gasb^Y$ for a finite
set $Y$. We carry out the computation in a finite volume which is 
chosen so that the distance from $Y$ to the boundary of the finite
volume is sufficiently large. We will study how large the 
finite volume should be in the next section.

\section{Errors }

In this section we study the sources of error in our
computations of the $f(X)$. There are two: the use of a finite volume to 
compute infinite-volume quantities and the truncation determined by 
the cutoff $\cb$ in section \ref{compute_free}.

\begin{figure}[tbh]
\begin{center}
\includegraphics{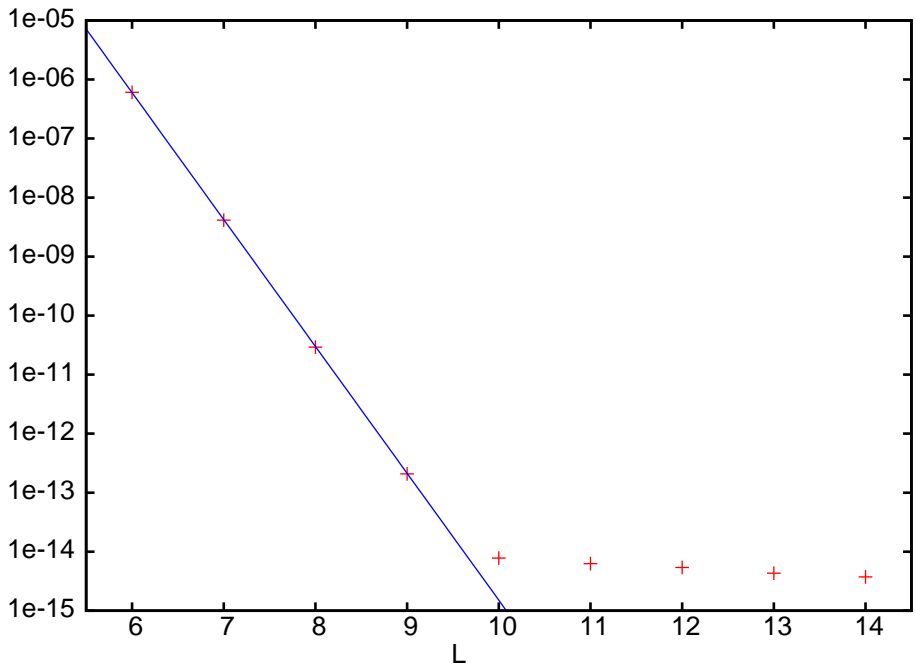}
\caption{\leftskip=25 pt \rightskip= 25 pt 
$L$ is a measure of the size of the finite volume. 
The quantity plotted is the average change in $f(X)$
when $L$ is decreased by $1$. 
(See eq. \reff{L_free}.) 
}
\label{rectangle}
\end{center}
\end{figure}

We choose the finite volume in which we carry out our 
calculation as follows. 
The block-spin configurations $\gasb$ that we consider are of 
the form $\gasb^Y$ for finite sets $Y$. We take these sets $Y$ 
to be centered near the origin and take the finite volume to 
be a square centered at the origin. 
The square contains the blocks with centers at 
$(2i,2j)$ with $-L \le i \le L$, $-L \le j \le L$. 
So the infinite-volume limit is obtained by taking $L \ra \infty$. 

To study the finite-volume error in our calculation we do the following. 
The free energy $f(Y)$ depends on $L$, so we denote it by $f_L(Y)$. 
As a measure of the finite-volume error we use
\bea
{1 \over N} \sum_Y | f_L(Y) - f_{L-1}(Y)|
\label{L_free}
\eea
where the sum is over one element of each translation class with 
$s(Y) \le 210$, and $N$ is the number of terms in the sum.
In this study of the finite-volume error we 
take $\cb=30$. This is much smaller than the cutoff we used for 
the main calculations, but the behavior of the finite-volume error 
with $L$ is the same whether we look at this smaller set of 
coefficients or the larger set.

This average difference as a function of $L$ is shown in figure
\ref{rectangle}. The vertical scale is logarithmic, so the approximately 
linear dependence seen for the smaller values of $L$ indicates exponential
decay of this difference with $L$. 
The line shown in the figure is of the form $c e^{-L/0.2}$. 
Keeping in mind that $L$ corresponds to numbers of blocks and 
the blocks are $2$ by $2$, the decay length of $0.2$ corresponds to 
a decay length of $0.4$ in units of lattice spacings. 
This very short decay length is a result of the block spin being $+1$ 
at all but a finite number of block sites.
Beginning with $L$ around 10 
the difference is dominated by numerical errors.
In our simulations we are very conservative and take $L=15$.
With this choice the finite-volume error is at the same level as the 
numerical error. 

\begin{figure}[tbh]
\begin{center}
\includegraphics{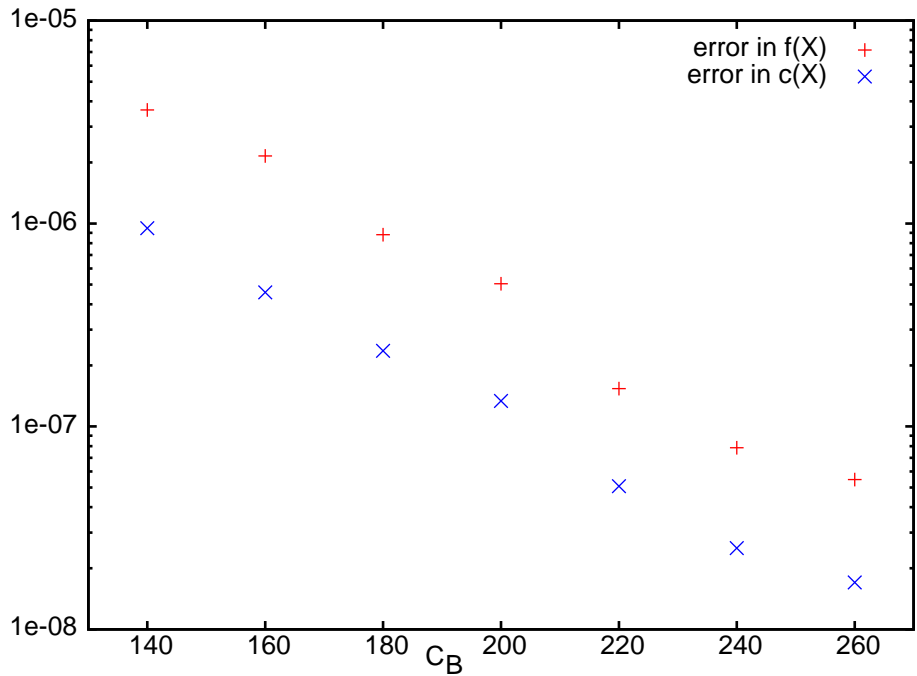}
\caption{\leftskip=25 pt \rightskip= 25 pt 
We plot the average of $f(X)-\bar{f}(X)$, as defined by eq. 
\reff{fbar}, and of $c(X)-\bar{c}(X)$, as defined by eq. \reff{cbar}, 
as a function of the cutoff $\cb$.
}
\label{error_rot_err}
\end{center}
\end{figure}

The translational symmetry of the original model implies that 
$f(X)$ is unchanged if we translate $X$. 
So we only need to compute $f(X)$ for one element of each translation class.
The model is also invariant under the dihedral group symmetry generated 
by rotations by mutiples of $\pi/2$ and reflections in the coordinate
axes. However, our method of computing $f(X)$ 
breaks the dihedral symmetry of the lattice, so 
our values of $f(X)$ for $X$'s from the same dihedral class 
are not exactly the same. The dihedral symmetry is only restored 
when we let $L \ra \infty$ and $\cb \ra \infty$. 
We have already seen that we can take $L$ sufficiently large 
that the finite-volume error is reduced to the order of the 
numerical error. So we can use the breaking of the dihedral 
symmetry to study how the error depends on the cutoff $\cb$. 

For various choices of $\cb$ we compute $f(X)$ for the same 
collection of $X$ as in our main calculation. 
Let $\bar{f}(X)$ be the average of $f(Y)$ over one $Y$ from each 
translation class which is related to $X$ by the dihedral 
symmetry. (The number of terms involved in this average ranges 
from $1$ to $8$.) The differences $f(X)-\bar{f}(X)$ are a measure 
of the amount of breaking of the dihedral symmetry and hence of the 
error in the computation from the cutoff $\cb$. We use 
the average
\bea
{1 \over N} \sum_Y | f(Y) - \bar{f}(Y)|
\label{fbar}
\eea
to quantify the error. 
As before the sum is over one element of each translation class with 
$s(Y) \le 210$, and $N$ is the number of terms in the sum.
This quantity is plotted in figure \ref{error_rot_err} as a function of $\cb$
for the free energies $f(Y)$. 
It is the higher set of points.
For the coefficients in the lattice-gas variables we define 
$\bar{c}(X)$ analogously, and study the average 
\bea
{1 \over N} \sum_Y | c(Y) - \bar{c}(Y)|
\label{cbar}
\eea
This quantity is the lower set of points in figure \ref{error_rot_err}.

\begin{figure}[tbh]
\begin{center}
\includegraphics{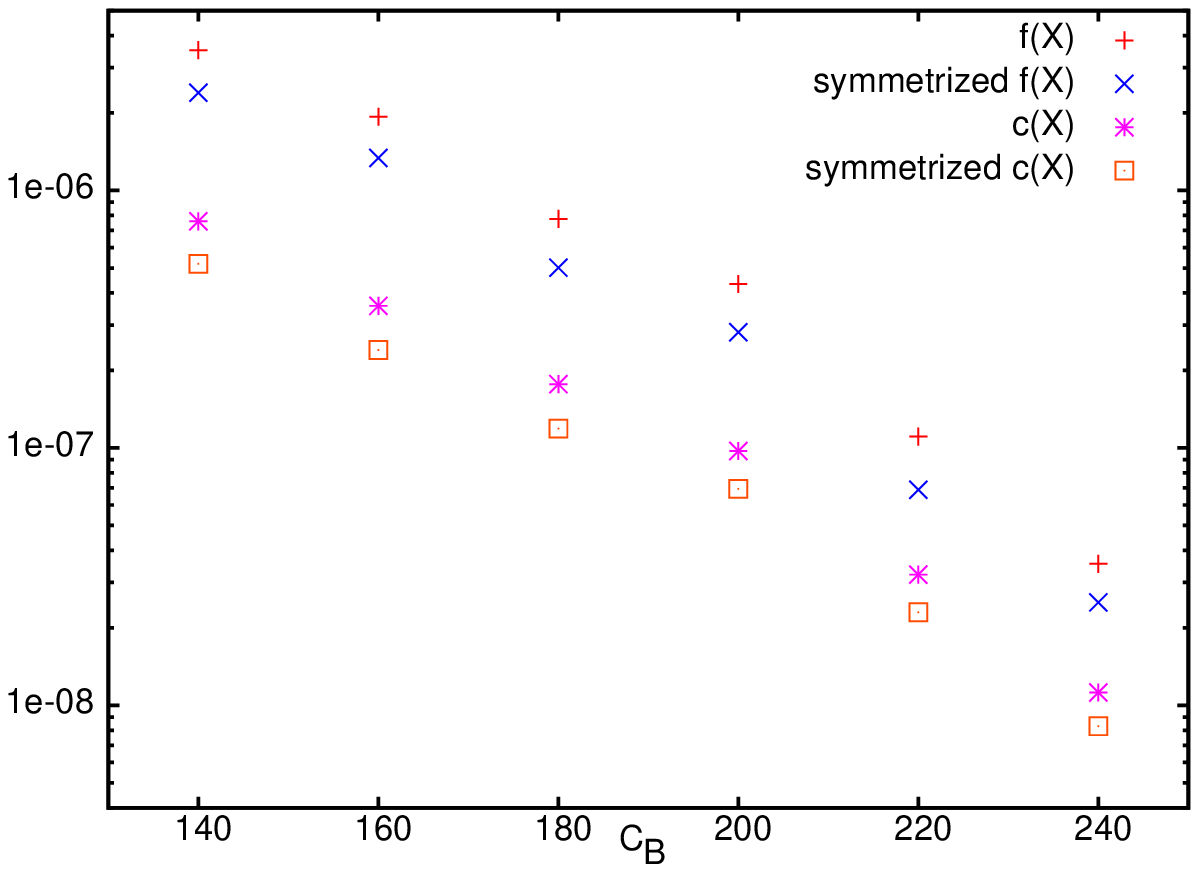}
\caption{\leftskip=25 pt \rightskip= 25 pt 
The convergence of four different quantities as $\cb \ra \infty$. 
From top to bottom the four quantities are given by equations 
\reff{quana} to \reff{quand}.
}
\label{error_dihedral_compare}
\end{center}
\end{figure}

We also study the convergence as $\cb \ra \infty$ in another way. 
Let $f^\infty(Y)$ denote $f(Y)$ for the largest value of $\cb$ which 
we use, i.e., 260. We then consider 
\bea
{1 \over N} \sum_Y | f(Y) - f^\infty(Y)|
\label{quana}
\eea
This is plotted as a function of $\cb$ in figure \ref{error_dihedral_compare}
for the free energy $f(Y)$. We also plot 
\bea
{1 \over N} \sum_Y | \bar{f}(Y) - \bar{f}^\infty(Y)|
\label{quanb}
\eea
As the figure shows, averaging over the dihedral group like this 
reduces the error somewhat.
The figure also includes the analogous plots for the coefficients in 
the lattice-gas representation, i.e., of the quantities  
\bea
{1 \over N} \sum_Y | c(Y) - c^\infty(Y)|
\label{quanc}
\eea
and
\bea
{1 \over N} \sum_Y | \bar{c}(Y) - \bar{c}^\infty(Y)|
\label{quand}
\eea

\section{Conclusions: }

We have shown that if we use lattice-gas variables, then the
computation of the coefficients in the renormalized Hamiltonian
only depends on a finite number of values of the renormalized 
Hamiltonian. So this computation does not depend on how we 
approximate the inherently infinite-dimensional renormalized 
Hamiltonian by a finite-dimensional approximation. 
We have also given a highly accurate method for computing the 
values of the renormalized Hamiltonian which takes advantage of 
the finite correlation length that results from the introduction 
of the renormalization group transformation.

The renormalized Hamiltonian has infinitely many different terms 
but the conventional wisdom is that it may be well approximated by 
a finite number of terms. In particular, the magnitude of the coefficients
should decay as the ``size'' of the set of lattice sites increases.
We studied this for the nearest neighbor critical Ising 
model on the square lattice under one step of the majority rule
renormalization group transformation. 
We computed a large number of coefficients in the lattice-gas
variables, ordered them by decreasing magnitude and plotted them.
We found that over several orders of magnitude 
the coefficients decayed exponentially with the number of terms, 
but the decay rate was slow. It takes about $850$ additional 
terms to see the magnitude reduced by just a factor of $1/2$.

If we use the usual spin variables, there is no natural way to 
compute the coefficients of the renormalized Hamiltonian.
We considered two methods of truncation.
If we look at an individual coefficient, we see significant dependence 
on the method used and on the value of the cutoffs used 
to specify the truncations in these methods.
Even with our computation of approximately $10,000$ values of 
the renormalized Hamiltonian, the uncertainty in the spin 
variable coefficients due to the different truncation methods 
is on the order of a percent for the largest coefficients and 
even larger as a percentage for some of the smaller coefficients.

One might hope to prove theorems about these 
real-space renormalization group transformations by defining a 
suitable Banach space of Hamiltonians and then doing a computer 
aided proof to show the transformation is defined in some open
subset of the Banach space and there is a fixed point in this subset.
Proving there is a fixed point would require constructing an approximation
to the fixed point with a finite number of terms.
Our numerical results suggest that at 
best such an approach will require a huge number of terms in the finite
approximation and at worst the number of terms needed will doom 
the approach to failure. 

Past numerical studies of the two dimensional Ising model using
the renormalization group have produced fairly accurate values 
of the critical exponents using a relatively modest number of 
terms in the renormalized Hamiltonian. 
These studies use the spin variables, so their accuracy is 
surprising given the difficulty we have found in computing the 
coefficients in the renormalized Hamiltonian accurately. 
An interesting question is to understand this.

\end{document}